\documentclass[runningheads]{llncs}
\usepackage{graphicx}
\usepackage{url}
\usepackage{color}
\usepackage[latin1]{inputenc}
\usepackage{subfig}
\usepackage{alltt}
\usepackage{amscd}
\usepackage{amsfonts}
\usepackage{amssymb}
\usepackage{caption}

\begin{document}

\pagestyle{headings}

\mainmatter

\title{ A Framework to build Games with a Purpose for Linked Data Refinement } 
\author{ Gloria {Re Calegari} \and Andrea {Fiano}  \and Irene {Celino} }
\institute{Cefriel, Viale Sarca 226, 20126 Milano, Italy
 \\ \email{\scriptsize{\{gloria.re,andrea.fiano,irene.celino\}@cefriel.com} }
}

\maketitle

\begin{abstract}
With the rise of linked data  and  knowledge graphs, the need becomes compelling to find suitable solutions to increase the coverage and correctness of datasets, to add missing knowledge and to identify and remove errors. Several approaches -- mostly relying on machine learning and NLP techniques -- have been proposed to address this \emph{refinement} goal; they usually need a \emph{partial gold standard}, i.e. some ``ground truth'' to train automatic models. Gold standards are manually constructed, either by involving domain experts or by adopting crowdsourcing and human computation solutions.

In this paper, we present an open source software framework to build \emph{Games with a Purpose for linked data refinement}, i.e. web applications to crowdsource partial ground truth, by motivating user participation through fun incentive. We detail the impact of this new resource by explaining the specific \emph{data linking} ``purposes'' supported by the framework (creation, ranking and validation of links) and by defining the respective \emph{crowdsourcing tasks} to achieve those goals. 

To show this resource's versatility, we describe a set of \emph{diverse applications} that we built on top of it; to demonstrate its reusability and extensibility potential, we provide references to detailed documentation, including an entire \emph{tutorial} which in a few hours guides new adopters to customize and adapt the framework to a new use case.
\end{abstract}

\section{Introduction}\label{sec:intro}
In the era of data-driven technologies, evaluating and increasing the quality of data is an important topic. In the Semantic Web community, the emergence of the so-called knowledge graphs has been welcomed as a success of the linked data movement but, in the meantime, has raised a number of research challenges about their management, from creation to verification and correction. The term knowledge graph refinement~\cite{paulheim2017knowledge} has been used to indicate the process to increase the quality of knowledge graphs in terms of finding and removing errors or adding missing knowledge. The same refinement operation is a challenge also for any linked dataset of considerable size. 

Addressing at scale the linked data refinement problem requires a trade-off between purely-manual and purely-automatic methods, with approaches that, on the one hand, adopt human computation~\cite{quinn2011human} and crowdsourcing~\cite{howe2006rise} to collect manually labeled training data and, on the other hand, employ statistical or machine learning to build models and apply the refinement operation on a larger scale. Indeed, active learning~\cite{settles2012active} and other recent research approaches in the artificial intelligence area have put back the ``human-in-the-loop'' by creating mixed human-machine approaches.

In this paper, we present an open source and reusable resource to build human computing applications in the form of games with a purpose~\cite{von2008designing} aimed to solve linked data refinement tasks. The presented resource consists of both a software framework and a crowdsourcing approach, that can be customized and extended to address different data linking issues.

The paper is organized as follows: Section~\ref{sec:related} presents related work; Section~\ref{sec:datalink} defines the refinement purpose addressed by our framework and Section~\ref{sec:crowd} explains the crowdsourcing task; the software resource is presented in Section~\ref{sec:enabler} and some applications built on it are illustrated in Section~\ref{sec:appl}; since it is a recently released resource, to explain its potential customization and to simplify its adoption, we set up an entire tutorial, briefly introduced in Section~\ref{sec:tutorial}; the code of the framework and the tutorial are available on GitHub and documented on Apiary (links throughout the paper); Section~\ref{sec:concl} concludes the paper.

\section{Related work}\label{sec:related}
Data linking is rooted in the record linkage problem studied in the databases community since the 1960s~\cite{fellegi1969theory}; for this reason, in the Semantic Web community, the term is often used to name the problem of finding equivalent resources on the Web of linked data~\cite{ferrara2013data}; in this meaning, data linking is the process to create links that connect subject- and object-resources from two different datasets through a property that indicate a correspondence or an equivalence (e.g. \texttt{owl:sameAs}). 

We prefer to generalize the concept of \emph{data linking} extending it to the task of creating links in the form of triples, without limitation to specific types of resources or predicates, nor necessarily referring to linking across two different datasets or knowledge graphs (data linking can happen also within a single dataset or knowledge graphs). In this sense, data linking can be interpreted as a solution to a \emph{linked data refinement} issue, i.e. the process to create, update or correct links in a dataset. As defined in~\cite{paulheim2017knowledge} with respect to knowledge graphs, with data linking we do not consider the case of constructing a dataset or graph from scratch, but rather we assume an existing input dataset which needs to be improved by adding missing knowledge or identifying and correcting mistakes.

The Semantic Web community has long investigated the methods to address the data linking problem, by identifying  linked dataset quality assessment methodologies~\cite{zaveri2016quality} and by proposing manual, semi-automatic or automatic tools to implement refinement operations~\cite{furber2010using,gueret2012assessing}. 
The large majority of refinement approaches, especially on knowledge graphs in which scalable solutions are needed, are based on different statistical and machine learning techniques~\cite{paulheim2014improving,dong2014knowledge,sleeman2013type,sleeman2015topic}. 

Machine learning methods, however, need a partial \emph{gold standard} to train automated models; those training sets are usually created manually by experts: while this usually leads to higher quality trained models, it is also an expensive process, so those ``ground truth'' datasets are usually small. Involving humans at scale in an effective way is, on the other hand, the goal of crowdsourcing~\cite{howe2006rise} and human computation~\cite{quinn2011human}.  Indeed, microtask workers have been employed as a mean to perform manual quality assessment of linked data~\cite{acosta2013crowdsourcing,simperl2011crowdsourcing}.

Among the different human computation approaches, games with a purpose~\cite{von2008designing} have experienced a wide success, because of their ability to engage users through the incentive of fun. A GWAP is a gaming application that exploits players' actions in the game to solve some (hidden) tasks; users play the game for fun, but the ``collateral effect'' of their playing is that the comparison and aggregation of players' contributions are used to solve some problems, usually labelling, classification, ranking, clustering, etc. Also in the Semantic Web community, GWAPs have been adopted to solve a number of linked data management issues~\cite{siorpaes2008games,waitelonis2011whoknows,chamberlain2008phrase,thaler2011spotthelink,hees2011betterrelations,celino2012linking,celino2012urbanopoly,wieser2013artigo,celino2014effectiveness}, from multimedia labelling to ontology alignment, from error detection to link ranking.

While the general guidelines and rules to build a GWAP have been described and formalized~\cite{law2011human} (game mechanics like double player, answer aggregation like output agreement, task design, etc.), building a GWAP for linked data management still requires time and effort. To the best of our knowledge,  source code was made available only for the labelling game Artigo~\cite{wieser2013artigo}.

\section{Data Linking Purpose}\label{sec:datalink}
As explained in Section~\ref{sec:related}, with \emph{data linking} we refer to the general problem of creating links in the form of triples. In this section, we provide the basic definitions and we illustrate the cases that our framework supports as purpose of the games that can be built on it.

\subsection{Basic Definitions}
The following formal definitions will be used throughout the paper.

\paragraph{\textbf{Resources}}
\textit{$\mathcal{R}$ is the set of all resources (and literals), whenever possible also described by the respective types. More specifically: $\mathcal{R} = \mathcal{R}_{s} \cup \mathcal{R}_{o}$, where $\mathcal{R}_{s}$ is the set of resources that can take the role of subject in a triple and $\mathcal{R}_{o}$ is the set of resources that can take the role of object in a triple; as said above the two sets are not necessarily disjoint, i.e. it can happen that $\mathcal{R}_{s} \cap \mathcal{R}_{o} \neq \varnothing$.}
 
\paragraph{\textbf{Predicates}}
\textit{	$\mathcal{P}$ is the set of all predicates, whenever possible also described by the respective domain and range.}
 
\paragraph{\textbf{Links}}
	\textit{$\mathcal{L}$ is the set of all links; since links are triples created between resources and predicates it is: $\mathcal{L} \subset \mathcal{R}_{s} \times \mathcal{P} \times \mathcal{R}_{o}$; each link is defined as $l=(r_{s},p,r_{o}) \in \mathcal{L}$ with $r_{s}\in\mathcal{R}_{s}, p\in\mathcal{P}, r_{o}\in\mathcal{R}_{o}$. $\mathcal{L}$ is usually smaller than the full Cartesian product of $\mathcal{R}_{s}, \mathcal{P}, \mathcal{R}_{o}$, because in each link $(r_{s},p,r_{o})$ it must be true that $r_{s}\in domain(p)$ and $r_{o}\in range(p)$.
} 

\paragraph{\textbf{Link scores}}
	\textit{$\sigma$ is the score of a link, i.e. a value indicating the confidence  on the truth value of the link; usually $\sigma \in [0,1]$; each link $l\in\mathcal{L}$ can have an associated score.}\\

\noindent One final note on subject resources: in the following sections, as well as in the framework implementation, we always assume that any subject entity can be shown to players in the game through some \emph{visual representation}; if the entity is a multimedia element (image, video, audio resource) this requirement is automatically satisfied; in other cases, some additional information about the subject may be required: e.g., a place could be represented on a map, a person through his/her photo, a document with its textual content, etc.

\subsection{Data Linking Cases}\label{sec:linkingcase}
Given the previous definitions, we can split the general data linking problem in a set of more specific cases as follows.

\paragraph{\textbf{Link creation}}
\textit{A link $l$ is created: given $\mathcal{R} = \mathcal{R}_{s} \cup \mathcal{R}_{o}$ and $\mathcal{P}$, the link $l=(r_{s},p,r_{o}), r_{s}\in\mathcal{R}_{s}, p\in\mathcal{P}, r_{o}\in\mathcal{R}_{o}$ is created and added to $\mathcal{L}$.}
	
All three components of the link to be created exist, i.e. they are already included in the sets $\mathcal{R}$ and $\mathcal{P}$.
It is important to note that \emph{\textbf{classification}} can be seen as a special case of link creation in which, given a resource $r_{s}\in\mathcal{R}_{s}$ to be classified and the predicate $p\in\mathcal{P}$ indicating the relation between the resource and a set of possible categories $\{cat_1,cat_2,\ldots,cat_n\}\subset\mathcal{R}_{o}$, the resource $r_{o}\in\{cat_1,cat_2,\ldots,cat_n\}$ is selected to create the link $l=(r_{s},p,r_{o})$.\\
For example, this is the case of music classification: given a list of resources of type ``music tracks'' in $\mathcal{R}_s$, the predicate \texttt{mo:genre} $\in\mathcal{P}$ and a set of musical styles in $\mathcal{R}_o$, the task is to assign the music style to each track by creating the link $(track,genre,style)$.\\
The case of link creation in which new resources and/or predicates are added to $\mathcal{R}$ and/or $\mathcal{P}$ (e.g. free-text labelling of images) is currently not supported by our framework, but it could be one of its possible extensions.

\paragraph{\textbf{Link ranking}}
\textit{Given the set of links $\mathcal{L}$, a score $\sigma \in [0,1]$ is assigned to each link $l$. The score represents the probability of the link to be recognized as true. Links can be ordered on the basis of their score $\sigma$, thus obtaining a ranking. }

In other words, we consider a Bernoulli trial in which the experiment consists in evaluating the ``recognizability'' of a link and the outcome of the experiment is ``success'' when the link is recognized and ``failure'' when the link is not recognized. Under the hypothesis that the probability of success is the same every time the experiment is conducted, the score $\sigma$ of a link $l$ is the estimation for the binomial proportion in the Bernoulli trial. \\
In the case of human computation, crowdsourcing or citizen science, each trial consists of a human user that evaluates the link and states that, in his/her opinion, the link is true (success) or false (failure); the human evaluators, if suitably selected, can be considered a random sample of the population of all humans; therefore, aggregating the results of the evaluations in the sample, we can estimate the truth value of a link for the entire population, by computing the probability of each link to be recognized as true. Then, ordering links on the basis of their score means having a metrics to compare different links on their respective ``recognizability''.\\
For example, this could be the case of ranking photos depicting a specific person (e.g. an actor, a singer, a politician): given a set of images of the person, human-based evaluation could be employed to identify the pictures in which the person is more recognizable or more clearly depicted.

\paragraph{\textbf{Link validation}}
\textit{Given the set of links $\mathcal{L}$, a score $\sigma \in [0,1]$ is assigned to each link $l$. The score represents the actual truth value of the link. A threshold $t\in[0,1]$ is set so that all links with score $\sigma\geq t$ are considered true.} 
	 
The difference between link validation and the previous case of link ranking is twofold: first, in link validation each link is considered separately, while in link ranking the objective is to compare links; secondly, while in link ranking  human judgment is used to estimate the subjective opinion of the human population, in the case of link validation the hypothesis is that, if a link is recognized as true by the population of humans (or by a sample of that population), this is a good estimation of the actual truth value of the link. The latter is also the reason for the introduction of the threshold $t$: while the truth value is binary (0=false, 1=true), human validation is more fuzzy, with ``blurry'' boundaries; the optimal value for the threshold is very domain- and application-dependent and it is usually empirically estimated.\\
An example of link validation would be assessing the correct music style  in audio tracks: it is well-known that sometimes music genres overlap and identifying a music style could also be subjective (e.g. there is no strict definition of what ``rock'' is); employing humans in this validation would mean attributing the most shared evaluation of a music track's genre.

As mentioned before, in the last two cases, the human evaluation of a link can be considered a Bernoulli trial: each link $l$ is assessed $n$ times by $n$ different users $u$; the link is recognized as true $X$ times (with of course $X\leq n$); each user $u_i$ can be more or less reliable and, in some cases, it is possible to estimate his/her reliability $\rho_{i}$. Therefore, the score of a link is $\sigma = f(n,X,\rho)$, i.e. it is a function of the number of trials $n$, the number of successes $X$ and the reliability values of the involved users $\rho=\{\rho_{1},\rho_{2},\ldots,\rho_{n}\}$.

\section{Crowdsourcing tasks for Data Linking}\label{sec:crowd}
Our framework allows to design and develop GWAP applications to solve data linking issue. In other words, the games built on top of our framework ask players to solve atomic data linking issues as basic tasks within the gameplay.

It is worth noting that building a GWAP does not automatically guarantee to collect enough players/played games to solve the data linking problem at hand; however, in our experience, if the task to be solved is properly embedded in a simple game mechanics and if the game is targeted to a specific community of interest, a GWAP is a valuable means to collect a ``ground truth'' dataset to train machine learning algorithms~\cite{re2018human}.

\subsection{Game basics}
Each GWAP built with our framework is a simple casual game organized in rounds; each round is formed by several levels and each level requires the player to perform a single action, which corresponds to the creation, ranking or validation of a  link. 
According to the definition of von Ahn~\cite{von2008designing}, each GWAP is an \emph{output-agreement double-player game}: users play in random pairs and the game score is based on the agreement between the answers provided by the players (i.e. if they agree, they get points). Our framework does not require users to play simultaneously, because it implements a common strategy in this kind of games, in which a user plays with a ``recorded player'', so the game scoring is obtained by matching answers provided at different times.

Our framework allows for both time-limited game rounds, in which players can answer to a variable number of tasks per round depending on their speed, and for level-limited rounds, in which players have a maximum number of tasks to address in each round; the choice of either option depends on considerations related to the specific task difficulty and to the game incentive mechanism.

The game adopts a \emph{repeated linking} approach by asking different players to address the same data linking task; conversely, the same task is never given twice to the same player. The ``true'' solution of a data linking task, therefore, comes from the aggregation of the answers provided by all the users who ``played'' the task in any game round. The number of players required to solve a data linking task depends on the aggregation algorithm as explained in the following. 

It is worth noting that the game scoring (i.e. points gained by players) is not directly related to the data linking scoring (i.e. the attribution of a score $\sigma$ to a link $l$): the former is an engagement mechanism to retain players, the latter is the very purpose of the game. 

\subsection{Atomic tasks and truth inference}
As mentioned above, the atomic task in any GWAP built with our framework is an individual data linking task. For example, in the case of music classification, a player could be given an audio track (the resource $r_s$), the relation \texttt{mo:genre} (the predicate $p$) and some options for music genres (e.g., classical, pop, rock, electronic, representing the potential objects of the link $l$); the action for the player would be to choose the genre (the resource $r_o$, say `rock') that better describes the audio track. By performing this action (the atomic task), the player is saying that he believes that the link $l=(r_s,p,r_o)$ is ``true''; the game can therefore alter the truth score $\sigma$ of the link $l$ by incrementing it. 

Inspired by record linkage literature~\cite{fellegi1969theory}, we assume each link score to start from $0$ and to be incrementally increased at each user contribution. The aggregation algorithm (also known in literature as \emph{truth inference} algorithm) implemented in our framework computes the score $\sigma$ as follows:
	\[ \sigma =  \delta_{+} \cdot \sum_{i} \rho_{i}	\]
where $\delta_{+}$ is the increment and $\rho_{i}$ is the reliability of the $i$th player which provided a solution to the task. 

In record linkage, where scores are assigned at each possible couple of records, the ``matching'' score is increased while the ``non-matching'' scores are decreased respectively. This is also supported by our framework, which allows to decrease the score of the links evaluated as ``false'' by the players. In the example above, if the music genres were mutually exclusive, the game could also alter the scores $\bar{\sigma}$ of the links $\bar{l}=(r_s,p,r_{\bar{o}})$ where $r_{\bar{o}} \neq r_o$ by decrementing them. The more general aggregation formula then becomes:
	\[ \sigma =  \delta_{+} \cdot \sum_{i} \rho_{i} - \delta_{-} \cdot \sum_{j} \rho_{j}	\]
where  $\delta_{+}$ and $\delta_{-}$ are the increment and decrement quantities, and $\rho_{i}$ and $\rho_{j}$ the reliability of the players judging the link as true or false, respectively.

The reliability of the player introduced in the equations above represents the level of trustworthiness in the collected answers and is used to weight players' contributions in the truth inference process. Intuitively, the answers provided by ``reliable'' players should count more than those provided by users who simply give random answers. 

As common practice~\cite{ul2013effects}, we adopt assessment tasks to quantify reliability: in each game, we compute player reliability as a function of the number of errors the player makes on those assessment tasks. The reliability value is 1 if the player makes no mistakes and decreases toward zero with increasing errors. As proposed in~\cite{celino2012linking}, by default our framework computes the player local reliability $\rho$ as follows:
	\[ \rho=e^{-k \cdot m}	\]
where $m$ is the number of mistakes per game round and $k$ a suitable constant. 

When the score $\sigma$ of a link overcomes a defined threshold $t$, the respective data linking task is considered solved and it is removed from the game, i.e. no other player will be asked to solve the same task. A data linking solved task is then used again in the game as assessment task to measure player reliability.

Interested readers can find examples of the input tasks, player contributions and aggregated output results of the Night Knights game (cf. Section~\ref{nk}) at \url{https://github.com/STARS4ALL/Night-Knights-dataset}.

\subsection{Tuning parameters}\label{sub:param}
Clearly, the truth inference mechanism explained before requires setting a number of parameters. The most suitable values depend on a mix of domain-specific considerations and empirical evaluation. Hereafter, we give some basic guidance.
\begin{itemize}
	\item \textbf{Reliability parameter $k$}: how much to penalize a player on the basis of his mistakes on the assessment tasks; this parameter should be tuned based on (1) how many assessment tasks are given in each round and (2) how much the reliability should be decreased on user's mistakes. For example, if each round contains $q$ assessment tasks, $k$ could be empirically set to $3/q$, so that, with $q$ errors, reliability $r$ decreases to $e^{-3}\approx0.05$ (i.e. player's answers are weighted very low because the user mistook all assessment tasks).
	\item \textbf{Score increment $\delta_{+}$ (and decrement $\delta_{-}$)}: how much the score should be incremented/decremented at each player answer; this usually depends on the minimum number of distinct users desired for truth inference (as well as on the threshold, cf. below). For example, if we think that $n$ players are needed, $\delta_{+}$ could be set to $1/n$, so that exactly $n$ players with full reliability are sufficient to reach score $\sigma=1$. 
	\item \textbf{Threshold $t$}: the value of the score $\sigma$ that must be overcome to consider a link solved; this parameter is also very context-specific and could be set to $1$ or to a value close to it. Of course, this threshold depends also on the setting of the above parameters.
\end{itemize}

\section{The GWAP Enabler framework} \label{sec:enabler}
To give a better idea of our software framework, we give some details on its technical internals. The GWAP Enabler is released as open source with an Apache 2.0 license and is made available at \url{https://github.com/STARS4ALL/gwap-enabler}.

\subsection{Structure of the framework}
Our GWAP Enabler is a template Web application formed by some basic software building blocks, which provide the basic functionalities to build a linked data refinement game; by customizing the ``template'', any specific GWAP application can be built. The three main components are the User Interface (UI), the Application Programming Interface (API) and the Data Base (DB).

The main table of the database is named \texttt{resource\_has\_topic} and contains all the links $l=(r_{s},p,r_{o}) \in \mathcal{L}$. 
Each link has a score $\sigma \in [0,1]$ which is updated every time it is played by a user.
The subject $r_s$ and object $r_o$ resources of links are stored respectively in the tables named \texttt{resource} and \texttt{topic}. For example, referring to the music classification case illustrated in the previous section, the \texttt{resources} table should contain the audio tracks and the \texttt{topics} table should list all the possible music styles. 
These three tables together contain all the data linking problem information and they are initially filled accordingly to the purpose of the specific GWAP. 

In addition to those tables, the database contains further information to customize the game according to the desired behaviour: the \texttt{configuration} table to change the truth inference parameters (cf. Section~\ref{sub:param}) and the \texttt{badge} table to change the badges given as reward to players during the gameplay. 
The remaining tables manage internal game data such as users, rounds, leaderboard or logs and are automatically filled during the gameplay; as such, they do not need to be modified or filled: they can of course be freely adapted at developers' will, being aware that altering those tables will require also modifying the code. 
 
\begin{figure}[t] 
	\centering
	\includegraphics[width=0.8\textwidth]{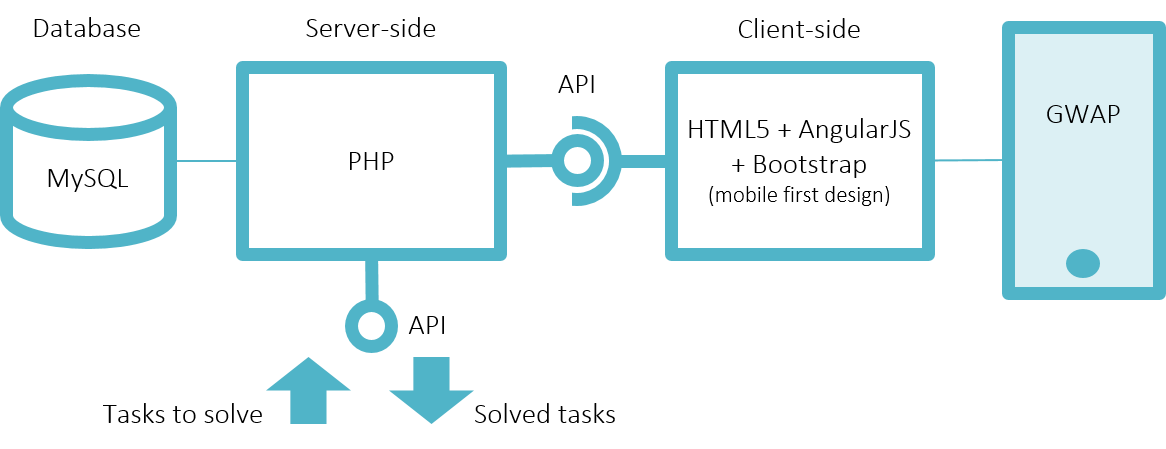}
	\caption {Architecture of the GWAP enabler}
	\label{fig:architecture}
\end{figure}

From a technical point of view, the GWAP Enabler is made up of an HTML5, AngularJS and Bootstrap frontend, a PHP backend to manage the logic of the application and a MySQL database, as shown in Figure~\ref{fig:architecture}. The communication between frontend and backend happens through an internal Web-based API, whose main operations consist in retrieving the set of links to be shown to players in each game round, updating the link score according to the agreement and disagreement of players and updating the points of the players to build leaderboards and assign badges.

Another external Web-based API is also set up to feed the game with new data and to access the GWAP results aggregated by the truth inference algorithm; in particular the API methods allow submitting new tasks to be solved, retrieving solved tasks (= refined links) in both JSON and JSON-LD formats, getting the number of tasks to be completed and listing the main KPIs of the GWAP (e.g., throughput, ALP and expected contribution~\cite{von2008designing}) for the game evaluation. This API is exhaustively documented at \url{https://gwapenablerapi.docs.apiary.io/}. 

Further details about the framework architecture can be found in the GitHub repository; the code of the enabler is in the ``App'' folder whereas the scripts to create the database are in the ``Db'' folder. Further details about the installation steps, the technical requirements, the database structure and some customization option are given in the wiki pages of the repository at \url{https://github.com/STARS4ALL/gwap-enabler/wiki}. 

\subsection{How to build a GWAP by using the framework}
To create a game instance out of the provided template, a developer should perform a series of operations and changes that affects the three building blocks constituting the framework.

First of all, the basic data linking case (cf. Section~\ref{sec:datalink}) and atomic crowdsourcing task (cf. Section~\ref{sec:crowd}) should be designed to address the specific use case. Then, the database has to be filled up with data by adding the core resources and links (\texttt{resource}, \texttt{topic} and \texttt{resource\_has\_topic} tables), the GWAP aggregation parameters (\texttt{configuration} table) and badges information (\texttt{badge} table). Please, note that a pre-processing phase is required to prepare the data and a careful analysis of the specific refinement purpose is an essential step to find and tune the proper parameters, thus this initial step could be long and complex depending on the context. 

Once data are in the DB, the code can be run as is or it can be tailored to address specific requirements; for example, game mechanics could be altered, further data to describe resources or links can be added (e.g., maps/videos/sounds) or a different badge/point assignment logic can be defined. 
Finally, the UI should be customized with the desired look and feel and the specific game elements (points, badges, leaderboards, etc.) could be modified to give the game a specific flavour or mood. 

\section{Existing applications built on top of the framework}\label{sec:appl}
We used the enabler to build three GWAPs that address the three different classes of data linking. Indomilando game aims to rank a set of cultural heritage assets of Milano, based on their recognizability; Land Cover Validation is an example of link validation game in which users are involved in checking land use information produced by two different (and disagreeing) sources; Night Knights is a game for both link creation and validation in which a set of images has to be classified into a predefined set of categories. 

\subsection{GWAP for link ranking: Indomilando}\label{indomilando}
\textit{Indomilando}\footnote{Cf. \url{http://smartculture-games.innovationengineering.eu/indomilando/}.}~\cite{celino2016analysis} is a web-based Game With a Purpose aimed to rank a quite heterogeneous set of images, depicting the cultural heritage assets of Milan. In each round the game shows the name of an asset and four photos, in which one represents the asset and the other three are put as distractors. The user has to choose the right picture related to the asset and, as an incentive, he gains points for each correct answer. A photo is removed from the game when it is correctly chosen three times. All the given answers on the photo (selection or non-selection of the picture) are recorded and analysed ex-post to measure ``how much'' the picture represents the asset: the intuition is that, the more a photo is correctly identified by players, the more recognizable it is. 

In Indomilando, we have a set of links $l$ that connects each photo with the asset it refers to; the assets and the photos are the subjects $r_s$ and objects $r_o$ of the links to be ranked. By counting and suitably weighting the number of times the pictures has been recognized (or not), we calculate the scores $\sigma$ of these links. Since they represent the probability of the links to be recognized as true, by ordering them we can rank the links, and thus the pictures of the cultural heritage assets of Milan, on the basis of their recognisability. 

The output of this game can be employed for various goals: selecting the best pictures representing an asset, understanding if an asset would benefit from further photo collection or evaluating if an asset may require additional promotional campaign because it is less recognized.

\begin{figure}[t]
	\centering
	{\includegraphics[width=0.45\textwidth]{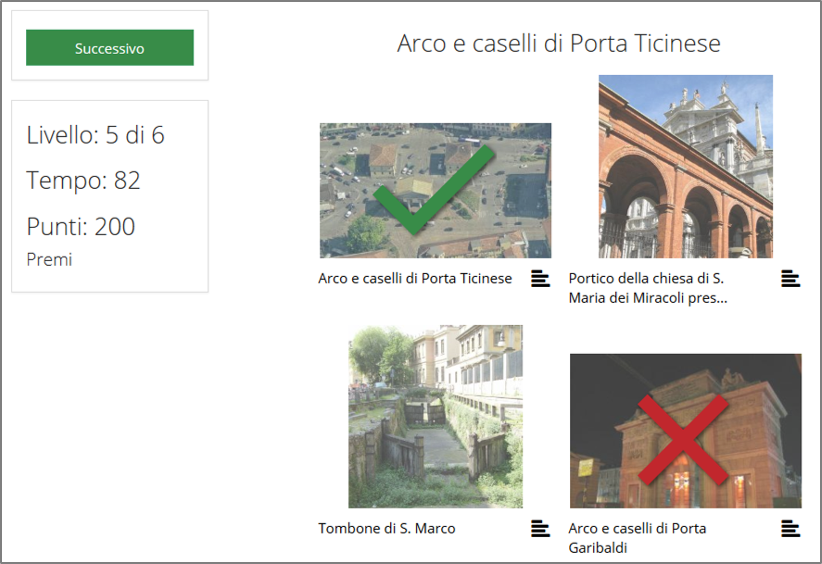}}%
	\quad
	{\includegraphics [width=0.50\textwidth]{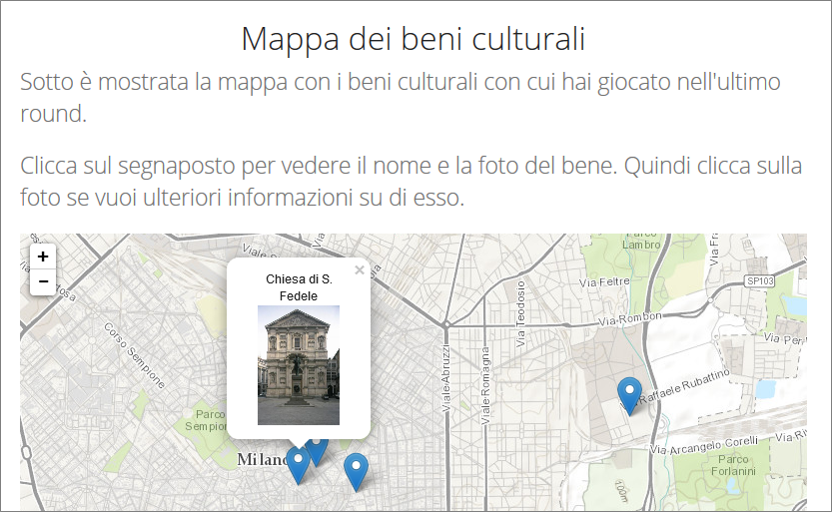}}%
	\caption {Indomilando: gameplay (left) and asset visualisation on a map (right).}
	\label{fig:indomilando}
\end{figure}

From a gamification point of view, users gain points for each correct answer and can challenge other players in a global leaderboard.  
Another incentive we give to players is the possibility to view on a map the assets they played with and to display their historic and cultural description, as shown in Figure~\ref{fig:indomilando}: this is an additional learning reward that Indomilando players showed to appreciate.
These incentives were very effective and the game had a great success: all the 2100 pictures we put in the game were played and ranked by 151 users, with a throughput of 125 photo ranked/hour.  

\subsection{GWAP for link validation: Land Cover Validation game}\label{lcvgame}
\begin{figure}[t] 
	\centering
	{\includegraphics[width=0.56\textwidth]{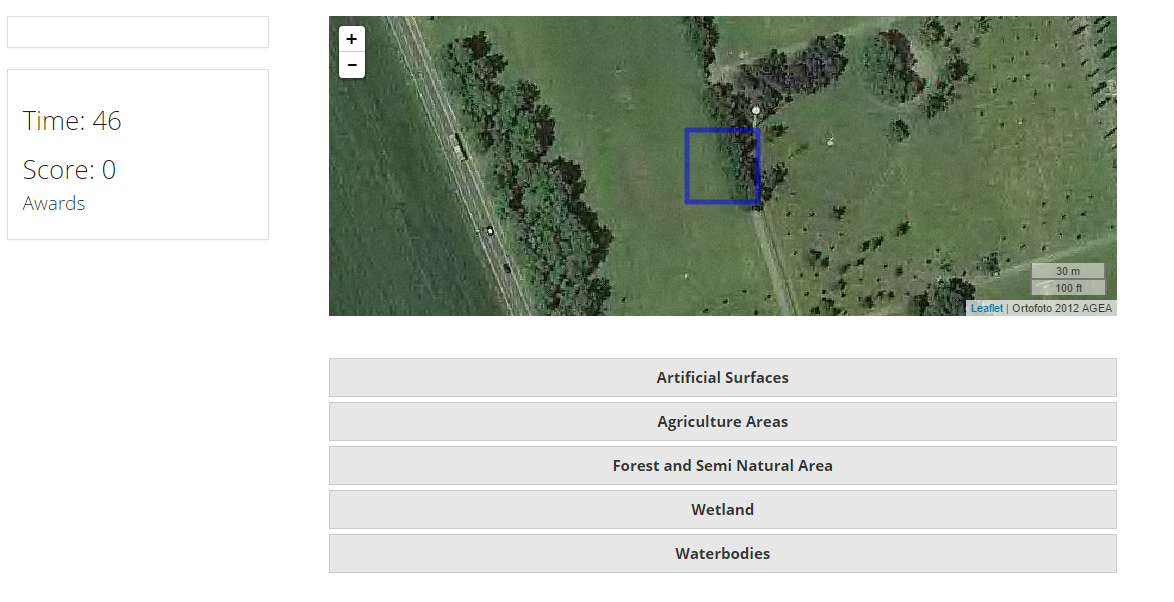}}%
	\quad
	{\includegraphics [width=0.41\textwidth]{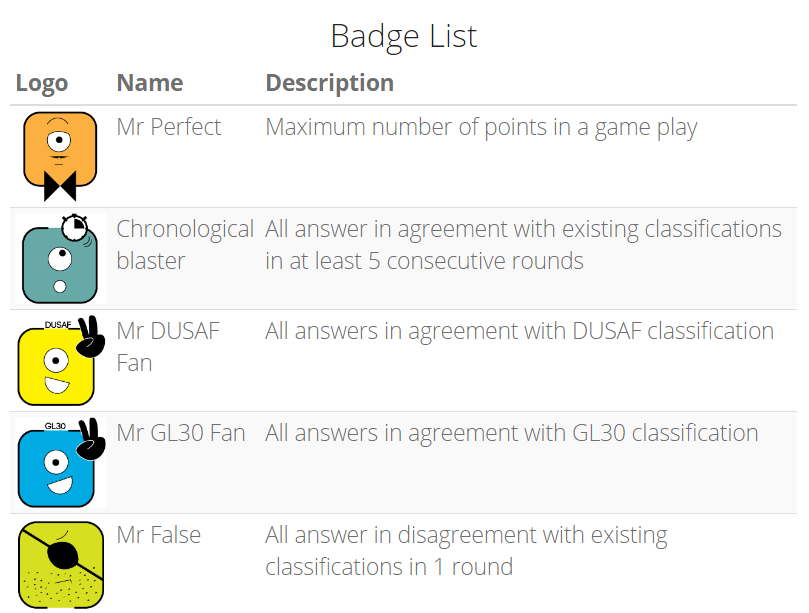}}%
	\caption {Land cover validation game:  pixel classification (left) and badges (right).}
	\label{fig:landcover}
\end{figure}

\textit{The Land Cover Validation game}\footnote{Cf. \url{http://landcover.como.polimi.it/landcover/}.}~\cite{brovelliland} is designed to engage Citizen Scientists in the validation of land cover data in the Como municipality area (in Lombardy, Italy). The player is requested to classify the areas in which two different land cover maps disagree: the DUSAF classification\footnote{Cf. \url{http://www.geoportale.regione.lombardia.it/en/home}.} made by Lombardy Region and GlobeLand30\footnote{Cf. \url{http://www.globallandcover.com/GLC30Download/index.aspx}.} provided by a Chinese agency. The validation is presented to the user as a classification task: given an aerial photo of a 30x30 square area (pixel), the player has to select the right category from a predefined list of land use types (i.e. residential, agricultural areas, etc.), as shown in Figure~\ref{fig:landcover}. As regards the incentives and the entertaining environment, players gains points and badges if they agree with one of the existing classifications and they can challenge other players in the leaderboard.  %\url{http://bit.ly/foss4game}.

From the data linking perspective, each pixel is the subject $r_s$ of two links, one connecting the pixel with its land cover defined by DUSAF ($r_{o_{DUSAF}}$) and the other with the GlobeLand30 classification ($r_{o_{GL30}}$). Each time one of the two land cover options is selected, the score $\sigma$ of the corresponding link is increased. This score represents the link truth value and a threshold is set so that all links with a score higher than this value are considered true. When a link score exceeds the threshold, the corresponding pixel is removed from the game. 

The game completely fulfilled its goal, since all the target 1600 aerial pixels were validated, thanks to 68 gamers that played more than 20 hours during the FOSS4G Europe Conference in 2015.

\subsection{GWAP for link creation and validation: Night Knights}\label{nk}
Night Knights\footnote{Cf. \url{http://www.nightknights.eu/}.}~\cite{re2018human} is a GWAP designed to classify images taken from the International Space Station (ISS) into a fixed set of six categories, developed within a project that aims to increase the awareness about the light pollution problem.

\begin{figure}[t]
	\centering
	\subfloat[Classify an image]{\label{fig:game}\includegraphics[width=.305\textwidth]{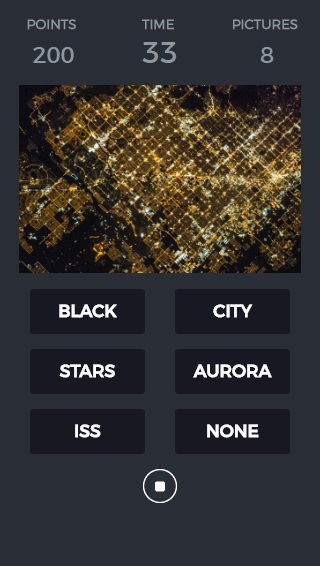}}%
	\quad
	\subfloat[Agreement]{\label{fig:agreement}\includegraphics[width=.300\textwidth]{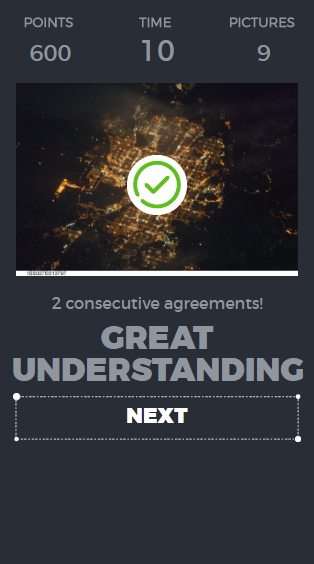}}%
	\quad
	\subfloat[Disagreement]{\label{fig:disagreement}\includegraphics[width=.303\textwidth]{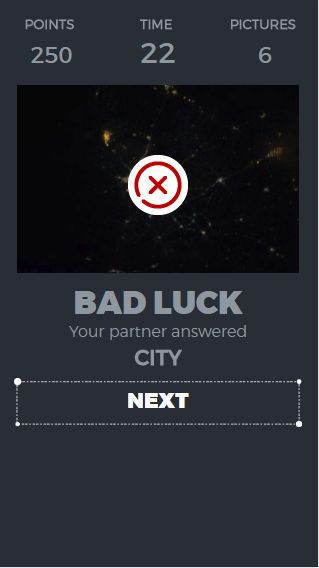}}%
	\caption{Night Knights: the game play}
	\label{fig:screenshot2}
\end{figure}

Each human participant plays the role of an astronaut that, coupled with another random player, has the mission of classifying as many images as possible playing one-minute rounds. As Figure~\ref{fig:screenshot2} shows, if players agree on the same classification they get points, which are collected to challenge other users in a global leaderboard.

The hidden goal of the game is to create new links between each image $r_s$ and its correct category $r_o$, by cross-validating them using the contributions of multiple users, suitably aggregated. The link creation algorithm works as follow: starting from a set of links connecting each image with all the available categories, the score $\sigma$ of a link is increased if a player chooses the corresponding category. Each image is offered to multiple players, whose contributions are weighted according to their reliability (measured with assessment tasks) and aggregated in the link score; once the score overcomes a specific threshold, the image is classified and removed from the game. By design, a minimum of four agreeing users are required to reach the classification threshold. 
More than 35,000 images have been classified since the launch of the game in February 2017 by 1,000+ players, with a throughput of 203 images classified/hour.

\section{A step-by-step tutorial to reuse the framework}\label{sec:tutorial} 
In the previous section, we showed how the GWAP Enabler was successfully employed to implement games to create and validate links through an image classification process and to rank links based on their recognizability. Since it is a new resource, in this section, we introduce a tutorial that  guides new adopters to customize and adapt the framework to a new use case. All the required changes to the GWAP Enabler sources are explained step-by-step in the GitHub repository at \url{https://github.com/STARS4ALL/gwap-enabler-tutorial}.

The goal of this tutorial is twofold; on the one hand, we provide developers with a guided example of an ``instantiation'' of our framework; on the other hand, we demonstrate how this GWAP Enabler could be used and adapted to build a  crowdsourcing web application to enrich and refine OpenStreetMap data (and, consequently, its linked data version LinkedGeoData), in which users are motivated to participate through fun incentives. 

More specifically, the GWAP application built in the tutorial is about classifying OpenStreetMap restaurants on the basis of their cuisine type. Data about restaurants are selected in a specific area (we give the example of the city of Milano, but developers can easily change it to their preferred location); those restaurants with an already specified type of cuisine are taken as ``ground truth'' (for the assessment tasks to compute player reliability), whereas all the remaining one are the subject resources, target of the classification process.

Players are randomly paired and are shown the restaurant name and position on a dynamic map; the game consists in finding an agreement on the cuisine type, selected in a set of predefined categories (the most widely used in OpenStreetMap). As a result, players contributions are aggregated in the truth inference process that implements the link collection and validation.

To create such an application, some changes to the framework's core functionalities are required; while in the GWAP Enabler by default each resource is displayed to players in the form of an image, in this tutorial scenario we want to show developers how to display both a textual information (the restaurant's name) and an interactive map (the restaurant's position). Therefore, this requires (1) to correctly store the relevant information in the database, (2) to modify the API code to retrieve the additional data and (3) to modify the UI code to display name and map. 

In the tutorial, we provide some basic instruction and we explain how to embed the map by using Leaflet\footnote{Cf. \url{http://leafletjs.com/}.}, an open-source JavaScript library for mobile-friendly interactive maps. We do not detail the graphical aspect, letting developers define their desired look-and-feel to give the game a more personal flavour or mood.  By going through the wiki instructions, developers can get their up and running GWAP in about half a day and they can gain enough knowledge about the framework to be able to reuse it for their own purpose, since the tutorial touches upon all the relevant modifications. 

\section{Discussion and Conclusion}\label{sec:concl}
In this paper, we presented an open source software framework to build Games With a Purpose embedding a crowdsourcing task for linked data refinement. The framework is aimed to help in the process of collecting manually-labelled gold standard data, which are needed as training set for automatic learning algorithms to implement refinement operations (link collection, ranking and validation) on large scale linked datasets and knowledge graphs. In other words, the presented framework helps to simplify the tedious and expensive human process of data collection, letting researchers focus on the subsequent steps of their scientific study and experimentation. 

We introduced the data linking cases implemented by the framework to explain its level of generality and potential for reuse; we illustrated the crowdsourcing task and truth inference process to clarify its design and possible customizations. Then, we gave some technical details about the internals of the GWAP Enabler, designed and developed accordingly to the most common Web development best practices, and we demonstrated the framework versatility by describing the diverse applications we built on top of it. Finally, we presented a step-by-step tutorial as a more detailed documentation and as a means to ease the reuse of this new resource; following the tutorial, a developer is guided to build an entirely new GWAP in a few hours, saving significant coding effort.

We provided all the references to get access to the framework code (released under an Apache 2.0 license) and its online documentation which consists of data schemas, API specification and sample input/output data, technical requirements and installation instructions, guided instruction to customize the GWAP Enabler.

The framework could be further extended to cover other refinement cases like free text labelling (i.e., insertion of new literal objects) or data linking issues related to the choice of different predicates.

\subsection*{Acknowledgments}
This work was partially supported by the STARS4ALL project (H2020-688135) co-funded by the European Commission. 

\bibliography{biblio}
\bibliographystyle{splncs04}

\end{document}